\documentclass[preprint,showpacs,preprintnumbers,amsmath,amssymb]{revtex4}

\usepackage{graphicx}
\usepackage{dcolumn}
\usepackage{bm}

\begin{document}

\preprint{}

\title{Monte Carlo approach of the islanding of polycrystalline thin films
}

\author{ F.Lallet, R.Bachelet, A.Dauger and N.Olivi-Tran} 
\affiliation{Laboratoire de Sciences des Proc\'ed\'es et Traitements de Surface, UMR-CNRS 6638, Ecole Nationale Sup\'erieure de C\'eramiques Industrielles, 47 avenue Albert Thomas, 87065 Limoges cedex,
France}

\date{today}

\begin{abstract}
We computed by a Monte Carlo method derived from the Solid on Solid model,
the evolution of a polycrystalline thin film
deposited on a substrate during thermal treatment. Two types of substrates have been studied:
a single crystalline substrate with no defects and a single crystalline
 substrate with defects. 
We obtain islands which are
either flat (i.e. with a height which does not overcome a given value)
or grow in height like narrow towers. 
A good agreement was found regarding the morphology of numerical nanoislands
at equilibrium, deduced from our model,
and experimental nanoislands resulting from the fragmentation of YSZ thin films
after thermal treatment.
\end{abstract}

\vfill
\maketitle
\section{Introduction}
In recent years, the formation of mesoscopic 
structures on crystal surfaces has become a subject of intense experimental
and theoretical study. 
Generally, for non periodically ordered nanostructures, the increasing specific 
area is favorable in order to enhance the physical properties (in optics,
semiconducting etc) owing to the increased number of active sites \cite{shalaev}.

We will study here the evolution of a thin film deposited on a single crystalline
substrate with or without defects. The thin film itself is polycrystalline
with the size of crystals corresponding to the thin film thickness.
The method employed experimentally to obtain such  thin films
is the sol gel processing \cite{brinker}.
The sol gel method proceeds as follows: a thin film is deposited on the substrate
by dip coating. After a first heat treatment (stage I), the thin film of nanometric
thickness is made of a large amount of nanocrystals of random orientation.
At this stage the film thickness is much larger than the mean size of these nanocrystals.
After a second heat treatment at higher temperature (stage II), thermal annealing induces grain growth. At this stage, the size of the crystals is of the order
of the film thickness. Simultaneously, the film is submitted to fragmentation
into more or less interconnected islands in order to reduce the total energy
and hence to reach a more stable state \cite{miller}.
 
The aim of this article if to model the formation of nanoislands
after thermal annealing of polycrystalline nanometric thin films
with no deposition.
Much literature has been written on models of the origin of islands in homoepitaxial
or heteroepitaxial single crystalline thin films \cite{shchukin,golovin,tekalign,spencer,spencer2,eisenberg,zhang}. Less model exist on nanoislands which
form spontaneously without deposition \cite{tekalign,olivi,olivi2}.
In this last case, surface roughening caused by the intrinsic elastic strain
and lattice mismatch between the thin film and the substrate has been widely
experimentally studied \cite{gao,jonsdottir}.
Theoretical studies showed that due to morphological variations in the shape of the surface
an originally flat film surface of a stressed solid thin film is unstable
\cite{spencer3,freund}. Experiments showed that film roughening
under various conditions and surface morphology form islands
 \cite{eaglesham,snyder,tersoff}, undulating surfaces \cite{yao,pidduck} 
and cusped surfaces \cite{jesson}.

Up to now most of the numerical and theoretical models (cited above) which represent
the evolution of a nanometric thin film are models which employ
a continuous representation of the elasticity of the thin film
and of the mismatches between the film and the substrate.
Moreover, only a  few articles deal with numerical model which
take into account polycrystalline thin films \cite{olivi,olivi2}.
The question is: is it possible to model the evolution of nanometric thin films
and nanometric islands with a continuous representation of their elasticity
and of the intrinsic strain resulting from lattice mismatches between the film and
the substrate?
To avoid this question, we used a Monte Carlo method applied to a non continuous
representation of a polycrystalline thin film. This model is derived
from the  Solid on Solid model but is applied here in the absence
of deposition. Our model is also derived from the two dimensional models
of polycrystalline materials which computed the evolution of polycrystalline
domains during thermal treatments \cite{srolo1,srolo2,srolo3}. Our model is based on
energetic considerations: we compute the energies resulting from the elastic
strains due to surface morphology of the thin film, the lattice mismatches and the grain boundary energies. We will see
that the resulting shapes of the islands depend on the relative
values of these three energies.

In section II, we shall present the model. In section III, results are shown
and discussed in the light of experimental results. Section IV corresponds to the conclusion.
 \section{Numerical procedure}
We modelled a thin polycrystalline film deposited on a
single crystalline substrate either perfect or
with a random distribution of defects.

Our model represents a thin film of $1nm$ thickness. Each domain
contains approximatively 500 to 1000 atoms.

In the case of a substrate with no defects, the domains have a square horizontal
section. The thin film is then represented by a square lattice of domains
with periodic boundary conditions. This model represents an polycrystalline
thin film deposited on a single crystalline substrate with no defects.

 In the case of a substrate with a random distribution of defects,
the domains have a section depending on the distribution of defects.
The defects are representative of the local maximum value of stress intrinsic to the substrate.
The locations of these stresses are obtained by generating a random array
of domains. Each domain is then corresponding to a local maximum value
of the stress in the substrate at the location of the domain.
The locations of the domains do not change during computation:
no displacements of the stress maxima are occurring in the substrate.
Hence the geometries of the domains correspond to the Voronoi
array of the distribution of defects. This model represents
a polycrystalline thin film deposited on a single crystalline substrate
with defects like dislocations, disinclinations and planar defects.

The mechanism of mass transport during thermal annealing
is surface diffusion:
\begin{equation}
(J(s+ds)-J(s))\Omega dt=\partial z \partial s
\end{equation}
where $J(s+ds)-J(s)$ is the transported number of atoms per unit time $t$
and $\partial z$ is the height difference in thin film thickness,
for changes in  the surface $s$ occurring during the mass transport of volume
$\Omega$
\cite{jullien}.

The flux $J$ may be also written \cite{jullien}:
\begin{equation}
J=-\frac{D_s \gamma \Omega^{1/3}}{k_BT} {\bf \nabla} k
\end{equation}
where $k$ is the surface curvature, $D_s$ is the surface diffusion constant, $\gamma $ is the surface tension,
$\Omega$ is the characteristic volume entering surface diffusion, $k_B$ is the Boltzmann constant
and $T$ is the absolute temperature.
In terms of the characteristic measures in the system we obtain:
\begin{equation}
J=-\frac{D_s \gamma \ell}{k_BT} \frac{1}{\Delta h}=\frac{D_s \gamma \ell}{\Delta h k_B T}
\end{equation}
where $\ell \propto \Omega^{1/3}$ is the characteristic mean size of the domains
 and $\Delta h$ is the difference of heights for two locations at a distance $\ell$.
Similarly, in terms of characteristic measures, equation (1) yields:
\begin{equation}
J=-\frac{\Delta h}{\ell \Delta t}
\end{equation}
for one time interval $\Delta t$ and for $s \propto \ell^2$.
Finally, relating equations (3) and (4) holds:
\begin{equation}
\Delta h=\sqrt{\frac{\ell^2 D_s \gamma \Delta t}{k_B T}}
\end{equation}
Hence, if we assume that the stress tensor inside the thin film is diagonal
(for example for cubic phase thin films) the energy related to $\Delta h$ using the work of  elastic forces, the Young modulus $Y$ and the Poisson ratio $\nu$ is:
\begin{equation}
E_h= Y (1+\nu)\Delta h \ell (h(t+\Delta t)-h(t))=  Y(1+\nu)\ell^2\sqrt{\frac{ D_s \gamma \Delta t}{k_B T}}(h(t+\Delta t)-h(t))
\end{equation}
where $(h(t+\Delta t)-h(t))$ is the displacement in the normal direction of the thin film for a time interval $\Delta t$ and $\ell$ is the displacement in
horizontal direction related to the Poisson ratio $\nu$.

If we deal with crystallographic orientations, let us report the behavior
of single crystalline grain growth \cite{srolo1} where the driving force
for change in the crystallographic orientation is related to the difference
in pressure:
\begin{equation}
\Delta p_{ij}=2 \gamma_b (\frac{1}{R_i}-\frac{1}{R_j})
\end{equation}
where $R_i$ is the sphere equivalent radius of the  domain $i$, $\gamma_b$ is 
grain boundary surface tension and $p_{ij}$ is the pressure. Several domains with the same crystallographic
orientation may form the same grain. 
Hence, the growth of one grain is equivalent to the changes in crystallographic
orientation of its neighbouring domains.
If
 we use the characteristic measures of the system, with  equation (7), we obtain the energy necessary to change the crystallographic orientation of one domain
by calculating the work of the driving force deduced from the pressure:
\begin{equation}
E_b= \gamma_b \ell^2 (\frac{1}{h}+\frac{1}{\ell})(c(t+\Delta t)-c(t))= \\ \gamma_b (\frac{\ell^2}{h}+\ell)(c(t+\Delta t)-c(t))
\end{equation}
where  $(c(t+\Delta t)-c(t))$ is the change in crystallographic orientation
in the dimension of a length, associated to the work of the driving force given by equation (7).

Straightforwardly, we consider here three aspects which contribute to the energy of our thin film
consisting of crystal species: the grain boundary energy (which is here
equivalent to the interfacial energy between two  elementary domains
of different crystallographic orientations), the interfacial energy
(which corresponds to the difference of energy between one elementary domain and the substrate) and  the surface energy (which is related 
here to the height of each elementary domain).
For our system of $N$ lattice domains, the energy necessary to change crystallographic orientation and height for domain $i$ with respect to domain $j$ becomes:
\begin{equation}
E_{ij}= B (\frac{\ell_i^2}{h_i}+ \ell) \sum_{j=1}^{NN}  (c_i-c_j) + \\
+C (\frac{\ell_i^2}{h_i}+\ell) \sum_{j=1}^{NN} (d_i-d_j)+ \\
D \ell_i^2 \sum_{j=1}^{NN}(h_i-h_j)
\end{equation}
here $h_i$ is the height of elementary domain $i$, $c_i$ is the coordinate of the crystallographic orientation in direct space on the horizontal plane, 
$d_i$ is the coordinate of the crystallographic
 orientation in direct space on the vertical plane,$\ell_i$ is the horizontal dimension  of domain $i$.
The first and second  terms of right hand side of the equality correspond
to the  interfacial energy coming from the grain boundary energy and from the energy with respect to the substrate.
And the third term corresponds to the  surface energy related to the heights of
 the domains (see equation (6)).

 $NN$ is the nearest neighbours number of a lattice domain. $B=\gamma_{b1}$ scales the interfacial energy between two
elementary domains and is the boundary surface tension. 
$C=\gamma_{b2}$ scales the interfacial energy with respect
 to the substrate where $\gamma_{b2}$ is the interfacial surface tension of the domains with respect to the substrate.
 $D=Y (1+\nu) \sqrt{\frac{ D_s \gamma \Delta t}{k_B T}}$ scales the surface energy obtained
for different heights of the elementary domains.
The values of $c$ and $d$ range from the upper value of orientation in horizontal and vertical planes respectively and have the dimension of a length.
The time interval $\Delta t$ is chosen to be constant.

For Monte Carlo simulations of single phase films, only one type of event, namely lattice domain reorientation, was considered \cite{srolo1,srolo2,srolo3}. In our model, the height
of each elementary domain is also submitted to changes like in the SOS model.
But, unlike traditional SOS models, a species at domain $i$ may change its orientation with respect of its nearest
neighbour and of the substrate.

In our model, each domain owns three states $(c,d,h)$.
$h$ has its value ranging from 0$nm$ to a value that depends
on the physical properties of the thin film as will be seen in the results.

To simulate the islanding of our thin film, prior to simulation,
all elementary domains were assumed to have a height of 1 $nm$
and a random  crystallographic orientation.
After such initialization, the Monte Carlo algorithm works according
to the classical Metropolis scheme \cite{metro}.  A lattice domain is chosen at random
for three  events (changes in the projections of the crystallographic orientation
and height exchange) occurring.
A neighbour of this domain is also chosen at random, and the energy given by equation (9) is computed. 

 The  probability for each event is given by $P$
in which $\Delta E=E_1-E_2$, where $E_1$ and $E_2$ are energies given by eq.(2), of the present configuration and the configuration which the system may reach 
respectively:
\begin{eqnarray}
P=1 if  \Delta E <= 0, \\
P=\exp( \frac{- \Delta E}{k_BT}) if \Delta E > 0
\end{eqnarray}
where $k_B$ is the Boltzmann constant and $T$ is the simulation
temperature.  Note that, as height exchange and crystallographic
reorientation are not independent events, it may occur
that a domain changes its height inducing a change in the orientational
energy of the domain.
Moreover, we used a Monte Carlo technique to study the statistical sampling
of the thin film geometry, due to its surface topology.
The values of $B$, $C$ and $D$ may vary as well as the absolute temperature
$T$.
\section{Results}
We computed all the results presented in the following figures
at a temperature $T=1800^oK$ which correspond to experimental data
(see below). The values of $c$ and $d$ are enclosed between 0.5 and 2.5. At $t=0MCS$ the numerical thin film is perfectly flat
and is $h=1nm$ thick i.e. all domains have the same height $h=1nm$.
Numerical results have been averaged over 5 runs. All the thin films
are represented by a square of edge equal to 100$nm$  divided in 10000 domains with periodic boundary conditions. For the square lattice,
each domain is 1$nm$ wide. For the random array, the  domains are randomly distributed.

In figure 1, top image, one can see the  resulting fragmentation of a thin film
deposited on a square lattice. This figure is obtained after $t=10^9MCS$.
The image on the bottom of figure 1 corresponds to a zirconia thin film
after thermal annealing at 1500$^oC$.
This experimental thin film has been deposited by a sol gel process
(introduced in section I) on a perfect single crystalline substrate
of $Al_2O_3$.

\subsection{Crystallographic orientations and heights
of the domains}
We performed the Monte Carlo process as written in section II,
for the two kinds of substrate i.e. with or without defects.

In figure 2a, one can see the evolution of the heights of the domains
as a function of MCS, for $B=1 J.m^{-2} J$,$ C=1 J.m^{-2}$ and $D=1 J.m^{-3}$, in the case
of a periodic (square) array of domains
Figure 2b corresponds to the same evolution with the same numerical
values of $B,C$ and $D$ but  for a random array of domains.
One may see as in these two figures that the number of domains
with heights $h=0nm$ increases until equilibrium is reached.
During the same time, the numbers of domains with heights $h=1,2,3$ and 4$nm$
increase
until reaching equilibrium at $t=10^5 MCS$.
There is a difference in the two figures for the number of domains
with height $h=0nm$: for a periodic array of domains this number is lower
than for a random array of domains. This allows us to say that defects
on the substrate lead to a larger dewetting of the thin film.
We will see below that, depending on the values of $B,C$ and $D$ this
dewetting may vary.

Figure 3a and 3b is the evolution of the orientations $c$ and $d$ as a function
of MCS, for $B=1 J.m^{-2}$,$ C=1 J.m^{-2}$ and $D=1J.m^{-3}$, in the case
of a periodic (square) array of domains, where $c$ and $d$ vary from 0.5 to 2.5.
Figure 3c and 3d corresponds to the evolution of the orientations $c$ and $d$ as a function
of MCS, for $B=1 J.m^{-2}$,$ C=1 J.m^{-2}$ and $D=1J.m^{-3}$, in the case
of a random array of domains and for the same range of $c$ and $d$.
In  figure 3a and 3b, we see that all domains (for a periodic array of domains)
 change from a random distribution
of crystallographic orientations to the heteroepitaxial crystallographic
orientation. Indeed, all domains for the two types of substrates
get the lowest crystallographic orientation regarding its energy
for $t=10^6MCS$ on.
In figure 3c and 3d we see that not all domains have reached the lowest
crystallographic orientation with respect to their energy when there are
defects on the substrate (for a random array of domains).

\subsection{Influence of factors $B,C$ and $D$}
In this section, we will study the influence of the numerical
values of $B,C$ and $D$ on the evolution of the heights of the domains
and of the orientations of the domains as a function of MCS.

In figure 4 the evolution of the crystallographic orientations $c$ and $d$ are plotted as a function
of MCS for $B=10^{-5}J.m^{-2}$, $C=1J.m^{-2}$ and $D=1J.m^{-3}$ for a square array of domains
(resp. for a random array of domains). This figure corresponds to an average over 5 runs.
The evolution of $c$ is clearly different for these values of $B,C$ and $D$ from the results obtained in figure 3, as well for a square array of domains
as for a random array of domains.

We may say that the influence of the value of the constant $C$ is the same
as for $B$ as these two constant are symmetrical and may be exchanged
in equation (9).

In figure 5 the evolution of the heights as a function
of MCS is shown for $B=10^5J.m^{-2}$, $C=10^5J.m^{-2}$ and $D=10^{-5} J.m^{-3}$ for a square array of domains (figure 5a and 5b)
and  for a random array of domains (figures 5c and 5d).
Figure 5 show the evolution of the number of domains with heights
ranging from $h=0nm$ to $h=10nm$ for the two kinds of substrates. The evolution
of the number of domains with heights ranging from $h=0nm$ to $h=4nm$
shows that dewetting is larger and faster. The numbers of domains of heights ranging from $h=5nm$ to $h=10nm$
 increase until reaching an equilibrium for $t=10^{9}MCS$.

\subsection{Comparison with experiments}
Yttria Stabilized Zirconia (YSZ) thin films were elaborated by sol-gel dip-coating.
 First of all, clear homogeneous sols were prepared from zirconium n-propoxide, 
acetylacetone and n-propanol. Yttrium nitrate ($Y(NO_3)_3(H_2O)_5$) dissolved 
in n-propanol was used as the $Y_2O_3$ precursor. The yttria content was set to 
10 mol so that YSZ is expected to crystallize in its cubic phase.
 A continuous amorphous film is realized by dip-coating after the setting 
of the zirconium n-propoxide concentration into the precursor solution 
equal to $[Zr] = 0.025 mol.l^{-1}$. The dipping speed was fixed to $1.67 mm.s^{-1}$.
 These parameters allow to control the thickness of the continuous films
 which is close to $5 nm $ in that case.
 The thickness was chosen to be small in order to maximize the interfacial
 effects. A primary thermal treatment at $600^oC$ induces the crystallization of
the films which is made of randomly oriented nanocrystals
 of zirconia. Finally, a last thermal treatment at high temperature
 (15$min$ at $1500^oC$) induces an abnormal grain growth driven by the interface
 that leads to the epitaxy of the thin films. Finally, a last thermal 
treatment at high temperature (15$min$ at $1500^oC$) induces the breaking up 
of the film and the formation of epitaxial YSZ islands.
YSZ is in the cubic phase.

The c-cut saphire substrate was roughly polished in order to create defects. The
 roughly mechanico-chemical polishing was realized with colloidal silica
 dispersed into an acid solution. A short thermal treatment at high temperature
 (set to 15$min$ at 1500$^oC$) was necessary to get rid of high residual polishing-induced 
strains and to perform a very small mosaicity allowing the epitaxy of the thin film.

The experimental values of the parameters are, for comparison with numerical
 results : $\ell=1nm$, $D_s=8.10^{-5}m^2.s^{-1}$ \cite{ridder},
 $\gamma_s=620.10^{-3}J.m^{-2}$ \cite{pascal}, $Y=300Pa$, $\nu=0.3$ and $k_BT=2J$ for 1000 atoms per domain and
$\Delta t$ is of the order of $10^4 s$. We can consider that, as YSZ is in the cubic phase at room temperature,
 the value of parameter $D$ does not change with orientation due to quasi isotropy. For example,
the data in literature only give an error of 20 \% on the value of $\gamma_s$
for all orientations \cite{pascal}. The values of $\gamma_{b1}$ and $\gamma_{b2}$
are of the same order as $\gamma_s$.

We compared a top view  of an experimental thin film
with a top view of a numerical thin film.
Figure 6 is the top view
of an experimental YSZ thin film on a c-cut sapphire substrate with defects.
These experimental defects have been obtained by rough mechanical-chemical
polishing.
Figure 7 is a top view of a fragmented numerical thin film on a substrate with defects
for the numerical values of $B,C,D$ corresponding to the experimental
 parameters. The grey disks correspond to islands of $1nm $ height, the dark squares
correspond to higher islands with a size inversely proportional to their height. 
\section{Discussion}
For the value $B=C=1J.m^{-2}$ and $D=1J.m^{-3}$, the resulting evolutions of either the crystallographic orientations
or the heights reach an equilibrium after $t=10^6 MCS$ (see figures 2 and 3). 

In the case of crystallographic orientations, almost all $c_i$ and $d_i$ reach the minimum value of the four values which existed at $t=0MCS$. This means that the thin film is heteroepitaxial
at equilibrium. More precisely, the vertical projection of the orientation in direct space reaches its minimum value with respect to the substrate and the horizontal
projection of the same orientation also tends to minimize its value with
respect to the orientation of its neighbouring domains.
This is a phenomenon of minimization of the energy due to crystallographic
mismatches: for small mismatches the energy is lower.

Let us analyze figure 4. In this case , $B=10^{-5}J.m^{-2}$ while
$C=1J.m^{-2}$ and $D=1J.m^{-3}$. The horizontal projection of the crystallographic orientations
in direct space 
depends on the initial configuration of the different horizontal
crystallographic orientations. As $B=10^{-5}J.m^{-2}$ is very low,
it induces a low
energy  corresponding to the crystallographic orientation. So depending on the relative numbers of
the different values of $c_i$ and as the value of the probability
of $c_i$ values exchanges is close to 1 for each MCS, the resulting
behavior of the number of relative values of $c_i$ follows
a random behavior. 
Equilibrium has not been reached in this case and the respective numbers
of the different values of $c$ may change even after $t=10^9 MCS$.
We checked this by making other runs of our program, and the resulting
evolution of $c$ was different for different averages . We obtained a textured thin film.

In the case of figure 5a corresponding
to the values $B=C=10^5J.m^{-2}$ and $D=10^{-5}J.m^{-3}$, the evolution of the number of domains with heights
$h=0,1,2,3,4nm$  for the square lattice
of domains follow the same typical behavior as for the case where
 $B=C=1J.m^{-2}$ and $D=1J.m^{-3}$ but with a faster dewetting. Figures 5b and 5d show the appearance of domains
with heights larger than 4$nm$.
In the case of the random array of domains, the dewetting process is strong at $t=10^9MCS$.
Moreover, one can observe in figures 5b and 5d that the numbers of domains
with heights $h=5,6,7,8nm$ is not negligible. 
These preceding numbers reach an equilibrium
after $t=10^9MCS$. In the case of figure 
5d, there are more domains which grow in height for a substrate with defects
compared to the perfect substrate giving the results of figure 5b.

Let us analyze the action of the substrate on the thin film. Our thin film
has the same domain distribution as the distribution of defects
on the substrate and can not change it. In the case of a random array
of domains, the numbers of neighbouring domains for sites located
on defects is larger; hence the probability for them to grow in height
is larger because exchanges of matter with neighbours are more numerous.
Indeed, for the case of a substrate with defects, the islands 
are preferentially located on the zones of the substrate where defects
are more numerous \cite{olivi}. This may be easily explained by
the Monte Carlo process: once a domain as been chosen at random,  crystallographic orientation and height exchange will occur preferentially with its nearest
neighbouring site.
Physically, this may be explained by the fact that the locations on the substrate 
where there are more defects (disinclinations, dislocations or planar defects)
 induce larger stresses in the thin film. Hence the minimization of energy
leads to the minimization of the stresses, by growing in height on these locations
and to the minimization of crystallographic orientations.
Domains with more numerous close neighbours will have the tendency
to grow in height. This phenomenon may be explained
by making a dimensional analysis of the energy given by equation (9):
this energy (in $J$) divided by the volume (in $m^3$) of one domain,
 leads to the stress (which is a force per surface unit
in $J.m^3$). If the domain has numerous neighbours, the energy and the stress 
will be large, hence the minimization of energy
leads to the release of the stress, by growing in height on these locations
and minimizing crystallographic orientations \cite{olivi}.

Less islands grow in height either for zirconia thin film deposited on perfect single crystalline substrates or for numerical thin film on a square lattice (see figure 1,  top for simulations and bottom for experiment).
In the case of a substrate with defects, comparison of the numerical thin film and the experimental thin film
after islanding occurred show a good agreement (see figure 6 and 7). Strong dewetting occurred
in both cases and given islands grow in height for substrates with defects.

\section{Conclusion}

We modelled the islanding, without deposition, of polycrystalline thin films by a Monte Carlo process
taking account of the crystallographic orientations of the grains and of the 
heights of each nanometric domain composing the underlying lattice representing
the thin film.
The governing equation allowing to compute the energy of each of these domains
takes into account the surface tension at free surfaces and at grain boundaries,the surface diffusion constant, the surface tension, and the elasticity of the thin film.
Depending on the values of these parameters, we obtain different evolutions
of the distribution of crystallographic orientations and of the dewetting.
The dewetting is larger for substrates with defects. This characteristic
is also obtained  in experimental thin films after thermal annealing.
Straightforwardly, for substrates with defects, experiments and numerical simulations
show islands which grow in height which is less the case on perfectly single
crystalline substrates.

\pagebreak

\begin{figure}
\includegraphics[width=14cm]{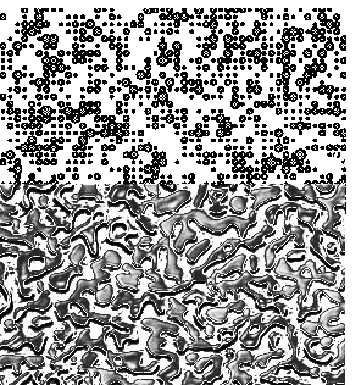}
\caption{Top: Fragmentation of a  numerical thin film with a square array of domains. The diameter of the circles is proportional to the height of the corresponding domain, Bottom: Electronic microscope image of a zirconia fragmented experimental thin film}
  \end{figure}
  
\begin{figure}
\includegraphics[width=14cm]{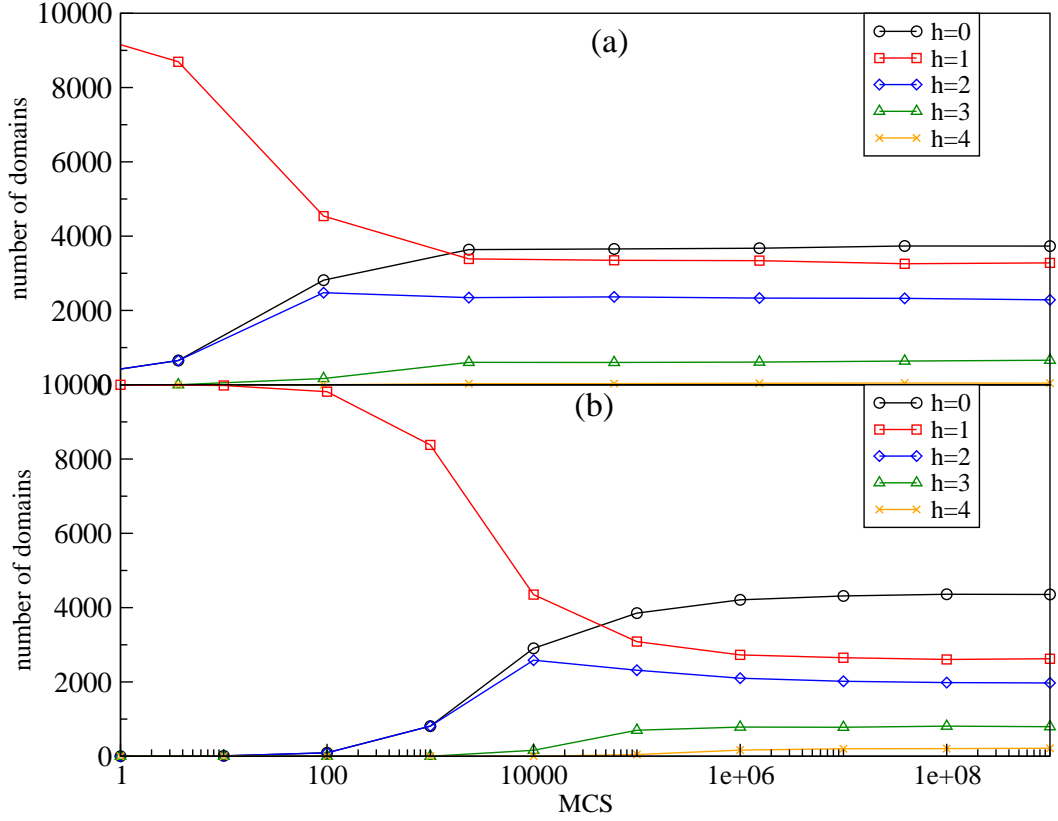}
\caption{(a) Evolution of the number of domains with heights $h=0,1,2,3,4nm$
as a function of MCS for a square array of domains corresponding to a substrate with no defects (b) Evolution of the number of domains with heights $h=0,1,2,3,4nm$
	as a function of MCS for a random array of domains corresponding to a substrate with  defects. The parameters for (a) and (b) are
 $B=1 J.m^{-2} J$,$ C=1 J.m^{-2}$ and $D=1 J.m^{-3}$ }
\end{figure} 

\begin{figure}
\includegraphics[width=14cm]{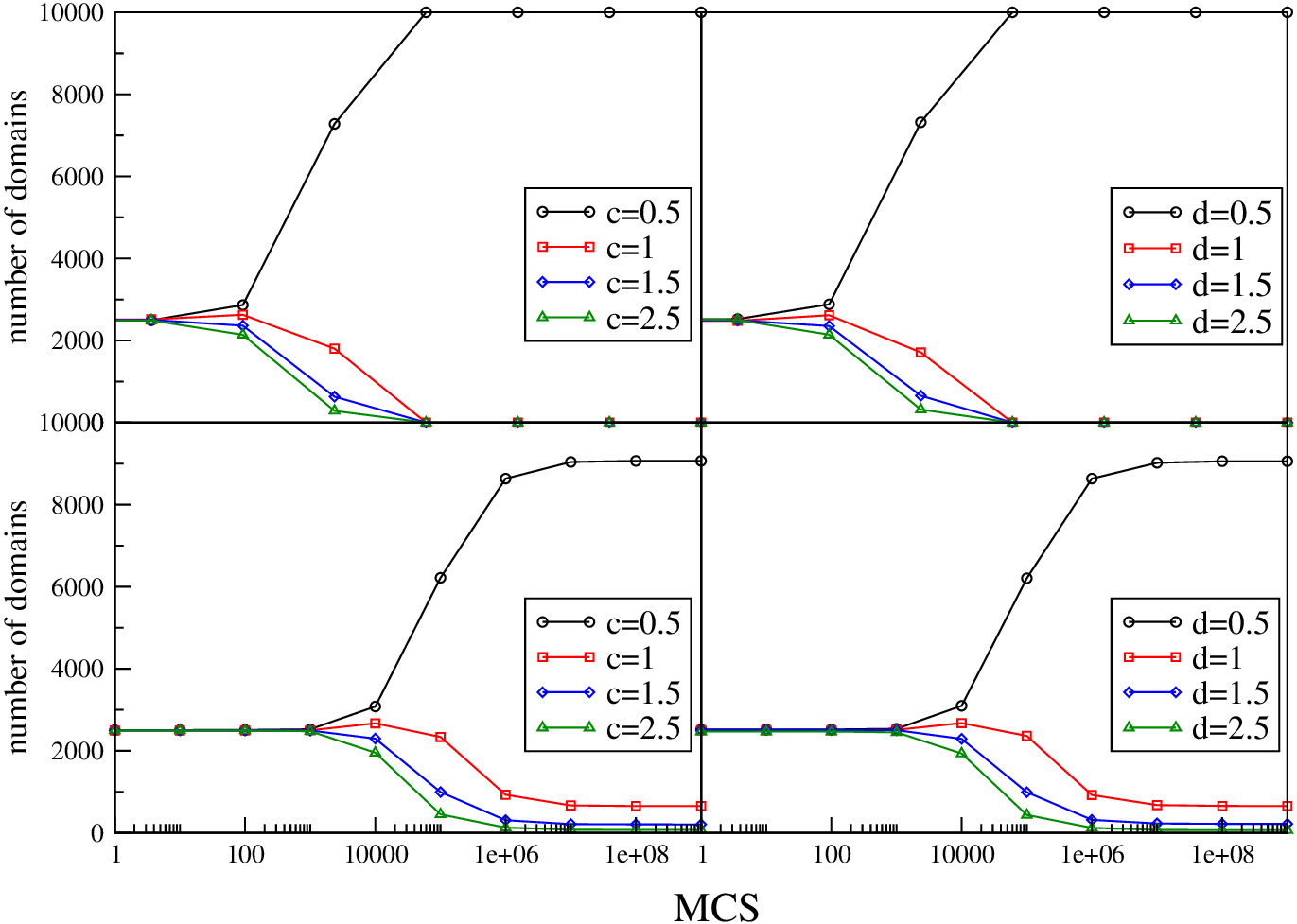}
\caption{(a) Evolution of the vertical crystallographic orientation $c=0.5;1;1.5;2.5nm$ 
as a function of MCS for a substrate with no defects (b)Evolution of the horizontal
 crystallographic orientation $d=0.5;1;1.5
;2.5nm$ as a function of MCS for a substrate with no defects (c)Evolution of the vertical crystallographic orientation $c=0.5;1;1.5
;2.5nm$
as a function of MCS for a substrate with  defects (d)Evolution of the horizontal
 crystallographic orientation $d=0.5;1;1.5
;2.5nm$ as a function of MCS for a substrate with  defects. The parameters
 for this figure are $B=1 J.m^{-2} J$,$ C=1 J.m^{-2}$ and $D=1 J.m^{-3}$
}
\end{figure}

\begin{figure} 
\includegraphics[width=14cm]{figure4.eps}
\caption{(a) Evolution of the vertical crystallographic orientation $c=0.5;1;1.5
;2.5nm$
as a function of MCS for a substrate with no defects (b)Evolution of the horizon
tal
 crystallographic orientation $d=0.5;1;1.5
;2.5nm$ as a function of MCS for a substrate with no defects (c)Evolution of the vertical crystallographic orientation $c=0.5;1;1.5
;2.5nm$
as a function of MCS for a substrate with  defects (d)Evolution of the horizontal
 crystallographic orientation $d=0.5;1;1.5
;2.5nm$ as a function of MCS for a substrate with  defects. The parameters for t
his figure are $B=10^{-5} J.m^{-2} J$,$ C=1 J.m^{-2}$ and $D=1 J.m^{-3}$
}
\end{figure}

\begin{figure}
\includegraphics[width=14cm]{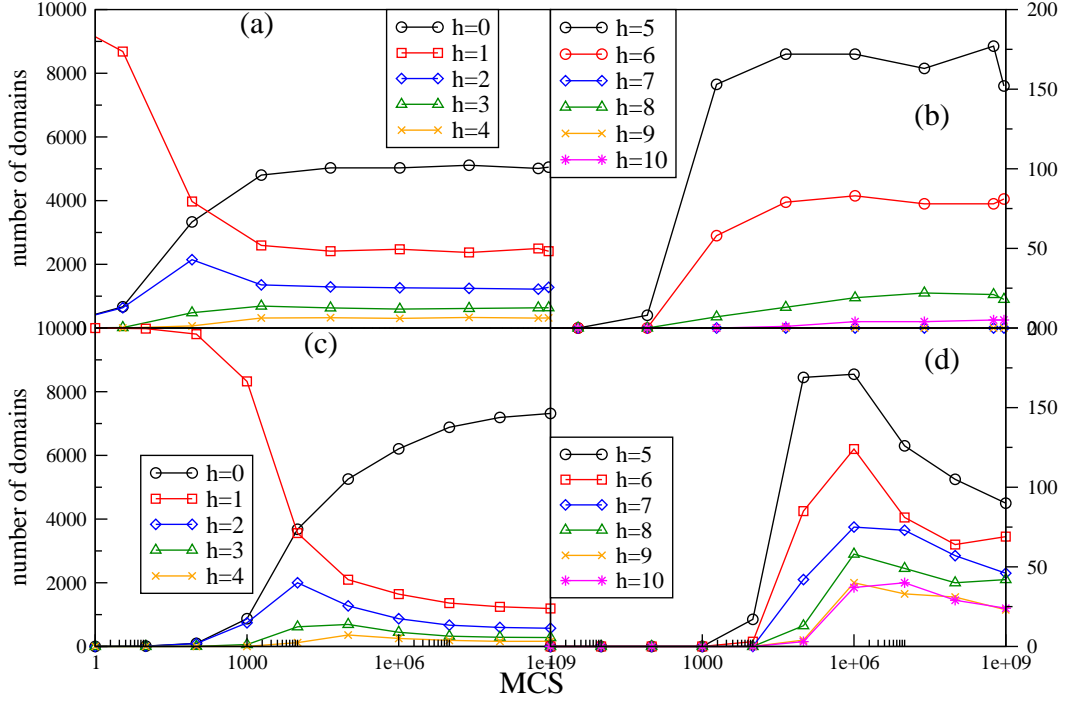}
\caption{(a) Evolution of the number of domains with heights $h=0,1,2,3,4nm$
as a function of MCS for a square array of domains corresponding to a substrate
with no defects (b) Evolution of the number of domains with heights $h=5,6,7,8,9,10nm$
as a function of MCS for a square array of domains corresponding to a substrate
with no defects(c) Evolution of the number of domains with heights $h=0,1,2,3,4
nm$
        as a function of MCS for a random array of domains corresponding to a substrate 
with  defects.(d) Evolution of the number of domains with heights $h=5,6,7,8,9,10
nm$
        as a function of MCS for a random array of domains corresponding to a substrate
with  defects. The parameters for (a),(b),(c) and (d) are $B=10^5 J.m^{-2} J$,$ C=10^5
 J.m^{-2}$ and $D=10^{-5} J.m^{-3}$}
\end{figure}

\begin{figure}
\includegraphics[width=14cm]{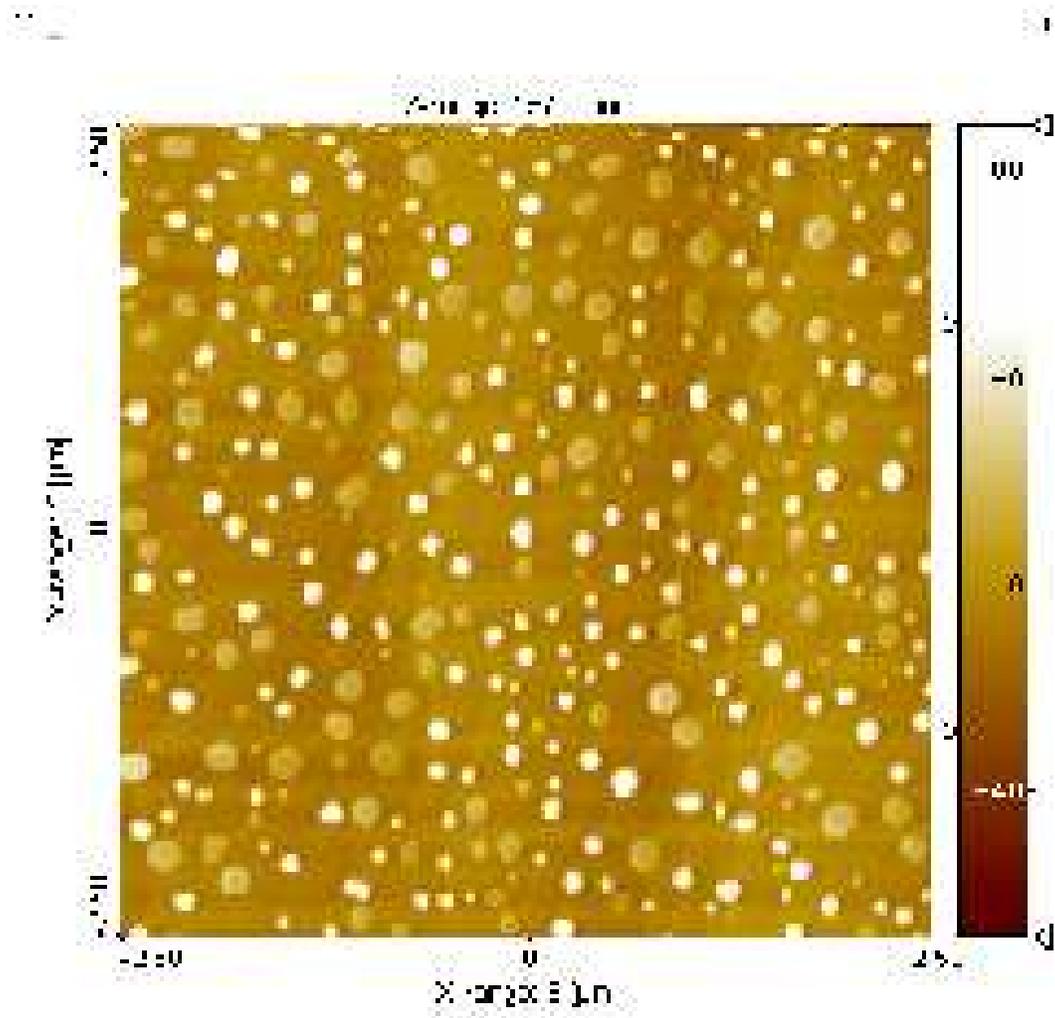}
\caption{AFM topography of a YSZ thin film islands on a substrate with defects}
\end{figure}

\begin{figure}
\includegraphics[width=14cm]{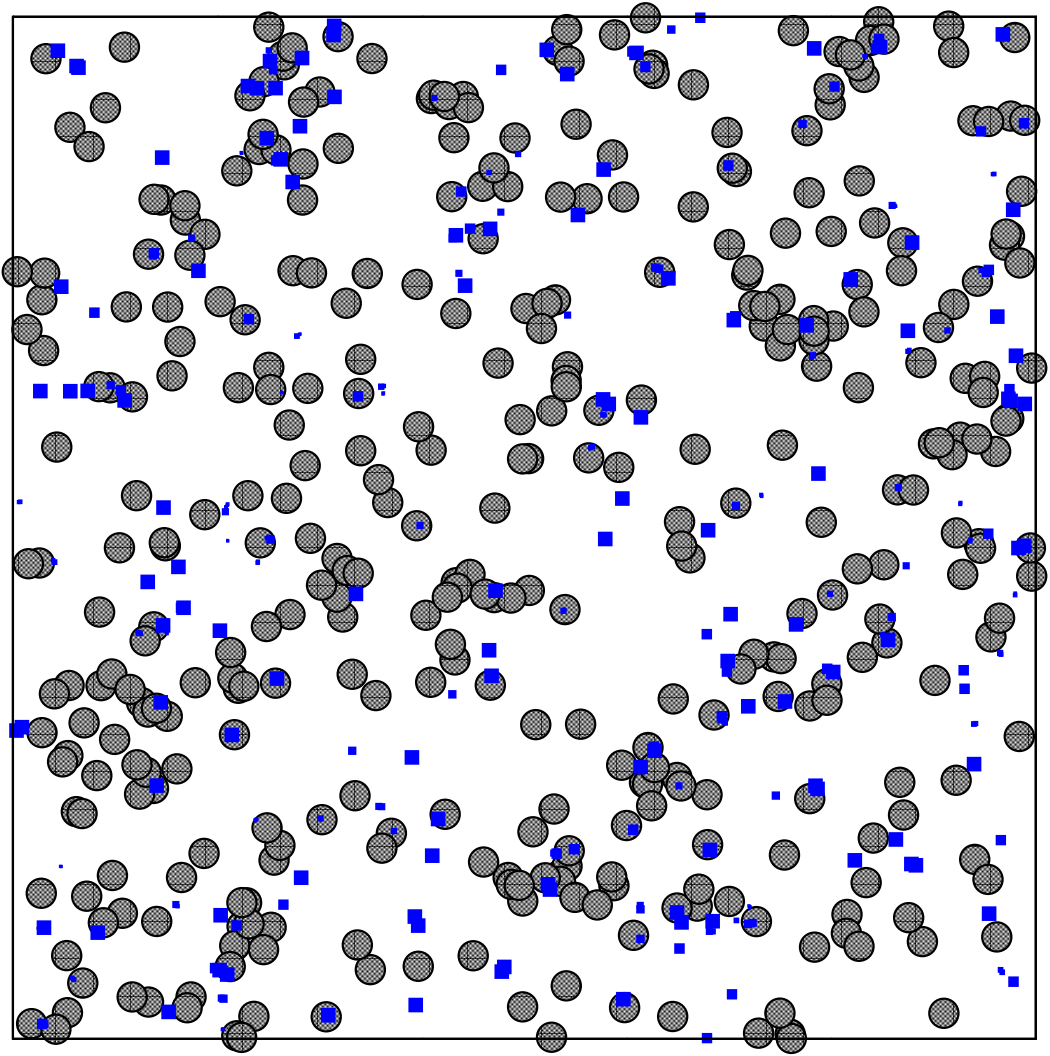}
\caption{Numerical image of islands corresponding to experimental values of parameters $B,C$ and $D$. Islands of height $h=1nm$ are in grey and islands of larger heights are in dark with a diameter inversely proportional to their height}
\end{figure}

\end{document}